\documentclass [12pt,a4paper]{article}
\usepackage{amssymb} 
\usepackage{graphicx}
\usepackage{graphics}
\usepackage{amsmath}
\usepackage{amsthm}
\newtheorem*{prop}{Proposition}

\newtheorem{lemma}{Lemma}

\def\proof{{\it Proof: }}

\def\qed{\nobreak\hfill $\square$}

\def\<{\langle}

\def\>{\rangle}

\def\iH{{\cal H}}

\def\Mm{M_m(\bbbc)}

\def\Mn{M_n(\bbbc)}

\def\M3{M_3(\bbbc)}

\def\M3r{M_3(\bbbr)}

\def\Mnsa{M_n^{sa}(\bbbc)}

\def\bbbr{{\mathbb R}}

\def\bbbc{{\mathbb C}}

\def\Tr{\mathrm{Tr}\,}

\newcommand{\R}{\mathbb{R}}

\newcommand{\C}{\mathbb{C}}

\newcommand*{\be}{\begin{equation}}

\newcommand*{\ee}{\end{equation}}

\newcommand{\tr}{\mathrm{Tr}}

\newcommand{\mc}[1]{\mathcal{#1}}

\newcommand{\abs}[1]{\left|#1\right|}

\newcommand*{\inner}[2]{\left<#1,\,#2\right>}
\newcommand*{\braket}[2]{\left<#1\right.\left|\,#2\right>}

\newcommand*{\bra}[1]{\left<#1\right|}
\newcommand*{\ket}[1]{\left|#1\right>}
\newcommand*{\ler}[1]{\left(#1\right)}
\newcommand{\Ln}{\mathrm{Ln}}

\newtheorem{Thm}{Theorem}

\theoremstyle{definition}
\newtheorem{ex}{Example}

\begin{document}
\vskip 1cm
\centerline{\LARGE {\bf Some inequalities for quantum Tsallis entropy}}
\bigskip
\centerline{\LARGE {\bf related to the strong subadditivity}}
\bigskip
\bigskip
\centerline{
{\bf D\'enes Petz}$^{a,}$\footnote{E-mail: petz@math.bme.hu} and
{\bf D\'aniel Virosztek}$^{b,}$\footnote{E-mail: virosz@math.bme.hu} }

\bigskip
\bigskip
\centerline{$^a$Alfr\'ed R\'enyi Institute of Mathematics}
\centerline{H-1364 Budapest, POB 127, Hungary}
\bigskip
\centerline{$^b$Budapest University of Technology and Economics}
\centerline{Egry J\'ozsef u.~1., Budapest, 1111 Hungary}
\bigskip
\bigskip


\begin{abstract}
In this paper we investigate the inequality
$S_q(\rho_{123})+S_q(\rho_2)\leq S_q(\rho_{12})+S_q(\rho_{23}) \, (*)$ where $\rho_{123}$
is a state on a finite dimensional Hilbert space $\iH_1\otimes \iH_2\otimes \iH_3,$
and $S_q$ is the Tsallis entropy. It is well-known that the strong subadditivity of
the von Neumnann entropy can be derived from the monotonicity of the  Umegaki relative
entropy. Now, we present an equivalent form of (*), which is an inequality of relative
quasi-entropies. We derive an inequality of the form
$S_q(\rho_{123})+S_q(\rho_2)\leq S_q(\rho_{12})+S_q(\rho_{23})+f_q(\rho_{123})$,
where $f_1(\rho_{123})=0$. Such a result can be considered as a generalization of the strong
subadditivity of the von Neumnann entropy. One can see that (*) does not hold in general (a picturesque example is included in this paper),
but we give a sufficient condition for this inequality, as well.
\end{abstract}


If $0 \le D \in B(\iH)$ is a state on a Hilbert space (or $0 \le D=\sum_i \lambda_i P_i
\in M_n(\bbbc)$ with $\Tr D=1$\,), then the von Neumann entropy is \be \label{Tsub4}
S(D)=-\Tr D \log\, D=-\sum_i \lambda_i \log \lambda_i\, \ge 0,
\ee
see in \cite{Bhat, H-P, O-PD}. If $D_{123}$ is a state on a Hilbert space $\iH_1\otimes \iH_2
\otimes \iH_3$, then it has reduced states $D_{12},\,D_2,\,D_{23} $ on the spaces $\iH_1
\otimes \iH_2, \, \iH_2$ and $\iH_2\otimes \iH_3,$ respectively, and the strong subadditivity is
$$
S(D_{123})+S(D_2) \le S(D_{12})+S(D_{23})\,.
$$
This result was made by E. Lieb and M. B. Ruskai in 1973 \cite{Lieb-Ruskai, O-PD}. Now we want
to make some extensions and the idea is $\ln_q \,x$. For any real $q,$ one can define the
deformed logarithm (or $q$-logarithm) function $\ln_q: \R_{+} \rightarrow \R$ by
$$
\ln_{q}\,x =\int_{1}^{x} t^{q-2} \mathrm{d}t=\begin{cases} \frac{x^{q-1}-1}{q-1} &\mbox{if }
q \neq 1 \, , \\ \ln\, x & \mbox{if } q=1\, . \end{cases}
$$
The corresponding entropy
$$
S_q(D)=-\Tr D \ln_q\, D
$$
is called Tsallis entropy \cite{A-D, Dar}. It is reasonable to restrict ourselves to
the $0<q$ case, because $\lim_{x \to 0^+} -x \ln_q \,x =0$ if and only if $0<q.$ If we
introduce the notation $\Ln_q\,x =-x \ln_q\,x$ we can write $S_q(D)=\Tr\,\Ln_q\,D$.\par

In this paper we present some inequalities on the Tsallis entropy which generalize or
are related to the strong subadditivity of the von Neumann entropy. We consider the
case of the classical probability theory, as well.


\section{The Tsallis entropy is subadditive, but not strongly subadditive} \label{bev}

If $D$ is a state on a Hilbert space $\iH_1\otimes \iH_2$, then it has
reduced states $D_1$ and $D_2$ on the spaces $\iH_1$ and $\iH_2$. The
subadditivity of the Tsallis entropy is 
\begin{equation}\label{Tsub}
S_q(D)\le S_q(D_1)+S_q(D_2),
\end{equation}
and it has been proved for $q>1$ by Audenaert in 2007 \cite{Au}.
In the $q>1$ case (\ref{Tsub}) can be written as
\begin{equation}\label{eq:sub2}
\Tr D_1^q+ \Tr D_2^q= ||D_1||_{q}^q+||D_2||_{q}^q\le 1+|| D||_{q}^q=1+\Tr D^q.
\end{equation}

The following example shows that in a special case the subadditivity can be verified by simple computations that are suggested by \cite{Furu}.

\begin{ex}
If $0 \le a, b, c, d$ and $a + b + c + d = 1$, then we can take the density
$D=\mathrm{Diag}(a, b, c, d),$ and the reduced densities are
$$
D_1=\mathrm{Diag}(a+b, c+d), \quad D_2=\mathrm{Diag}(a + c, b + d)\,.
$$

The inequality (\ref{Tsub}) is rather simple in this example.

For $1 \le q$ the above (\ref{Tsub}) is in this form:
$$
(a + b)^q + (c + d)^q + (a + c)^q + (b + d)^q \le 1 + a^q + b^q + c^q + d^q
$$
The case $q=2$ is rather trivial. For general $q,$ we use the function $f(x)=x-x^q$ ($x \in [0,1]$) and the identities $f(x)=x(1-x^{q-1})$
and $1-x^{q-1} y^{q-1}=1-x^{q-1}+x^{q-1}(1-y^{q-1})$. So
$$
f(a)+f(b)+f(c)+f(d)=
$$
$$
= a \ler{1-(a+c)^{q-1}\ler{\frac{a}{a+c}}^{q-1}}+
b \ler{1-(b+d)^{q-1}\ler{\frac{b}{b+d}}^{q-1}}
$$
$$
+c \ler{1-(a+c)^{q-1}\ler{\frac{c}{a+c}}^{q-1}}+
d \ler{1-(b+d)^{q-1}\ler{\frac{d}{b+d}}^{q-1}}
$$
$$
=(a+c)\ler{1-(a+c)^{q-1}}+(b+d)\ler{1-(b+d)^{q-1}}+
a(a+c)^{q-1}\ler{1-\ler{\frac{a}{a+c}}^{q-1}}
$$
$$
+c(a+c)^{q-1}\ler{1-\ler{\frac{c}{a+c}}^{q-1}}+
b(b+d)^{q-1}\ler{1-\ler{\frac{b}{b+d}}^{q-1}}
$$
$$
+d(b+d)^{q-1}\ler{1-\ler{\frac{d}{b+d}}^{q-1}}=
f(a+c)+f(b+d)
$$
\begin{equation} \label{szamol}
+(a+c)^q \ler{f\ler{\frac{a}{a+c}}+f\ler{\frac{c}{a+c}}}
+(b+d)^q \ler{f\ler{\frac{b}{b+d}}+f\ler{\frac{d}{b+d}}}.
\end{equation}

The concavity of $f$ gives
$$
(a+c)^q f\ler{\frac{a}{a+c}}+(b+d)^q f\ler{\frac{b}{b+d}} \leq (a+c) f\ler{\frac{a}{a+c}}+(b+d)
f\ler{\frac{b}{b+d}}\leq f(a+b)
$$
and
$$
(a+c)^q f\ler{\frac{c}{a+c}}+(b+d)^q f\ler{\frac{d}{b+d}} \leq (a+c) f\ler{\frac{c}{a+c}}+
(b+d)f\ler{\frac{d}{b+d}}\leq f(c+d).
$$
So we get
$$
(a+c)^q \ler{f\ler{\frac{a}{a+c}}+f\ler{\frac{c}{a+c}}}+(b+d)^q \ler{f\ler{\frac{b}{b+d}}
+f\ler{\frac{d}{b+d}}} \leq f(a+b)+f(c+d),
$$
hence from (\ref{szamol}) we get
$$
f(a)+f(b)+f(c)+f(d) \leq f(a+c)+f(b+d)+ f(a+b)+f(c+d).
$$
This is our statement. \qed
\end{ex}

The aim of this paper is to investigate the inequality
\be \label{Tub12}
S_q(\rho_{123})+S_q(\rho_2)\leq S_q(\rho_{12})+S_q(\rho_{23}),
\ee
where $\rho_{123}$ is a state on a Hilbert space $\iH_1\otimes \iH_2\otimes \iH_3$ (all
components are finite dimensional), and $\rho_{2}, \rho_{12}, \rho_{23}$ are the appropriate
reduced states. For $q=1,$ this is the well-known strong subadditivity (or with the usual
abbreviation: SSA) inequality (\ref{Tsub4}), which is a central result of the quantum
information theory \cite{Lieb-Ruskai}. \par

First, we have to note an easy consequence of Audenaert's theorem. The subadditivity implies that
$$
S_q(\rho_{123}) \leq S_q(\rho_{12})+S_q(\rho_3)
\text{ and } S_q(\rho_{123}) \leq S_q(\rho_1)+S_q(\rho_{23}).
$$
It follows that
$$
S_q(\rho_{123})+S_q(\rho_2)-S_q(\rho_{12})-S_q(\rho_{23}) \leq \mathrm{min} \{ S_q(\rho_1)+
S_q(\rho_2)-S_q(\rho_{12}),S_q(\rho_2)+S_q(\rho_3)-S_q(\rho_{23})\}.
$$
The Tsallis entropy is nonnegative and takes its maximum at the completely mixed state, and
the maximal value is $-\ln_q \frac{1}{d},$ where $d$ is the dimension of the underlying
Hilbert space. Therefore,
$$
S_q(\rho_{123})+S_q(\rho_2)-S_q(\rho_{12})-S_q(\rho_{23}) \leq -\ln_q \frac{1}{d_2}-\ln_q
\frac{1}{\mathrm{min}\{d_1, d_2\}},
$$
where $d_i$ is the dimension of $\iH_i.$ However, the strong subadditivity does not hold in general.

\begin{prop}
The only strongly subadditive Tsallis entropy is the von Neumann entropy, that is,  the strong subadditivity of the Tsallis entropy holds if and only if $q=1.$
\end{prop}

\proof
It is known that for product states the relation
\begin{equation} \label{classprod}
S_q(\rho_X \otimes \rho_Y)=S_q(\rho_X)+S_q(\rho_Y)+(1-q)S_q(\rho_X)S_q(\rho_Y)
\end{equation}
holds, hence the Tsallis enrtopy can not be subadditive for $q<1.$ In fact, it is neither
subadditive, nor superadditive \cite{Furu}. Therefore, the Tsallis enrtopy is not strongly
subadditive for $q<1.$ On the other hand, the next examples show that (\ref{Tub12})
does not hold for $1<q$. \par

Set $q>1$ and consider the matrix
$$
\rho_{123}=\left[ \begin {array}{cccccccc}  0 &0&0&0&0&0&0&0\\
\noalign{\medskip}0& 0 &0&0&0&0&0&0\\\noalign{\medskip}0&0& \frac{1}{4}&0& \frac{1}{4}&0&0&0\\
\noalign{\medskip}0&0&0& \frac{1}{4}&0& \frac{1}{4}&0&0\\\noalign{\medskip}0&0&
\frac{1}{4}&0& \frac{1}{4}&0&0&0\\
\noalign{\medskip}0&0&0& \frac{1}{4}&0& \frac{1}{4}&0&0\\\noalign{\medskip}0&0&0&0&0&0&0&0\\
\noalign{\medskip}0&0&0&0&0&0&0& 0
\end {array} \right].
$$
$\rho_{123}$ is positive and $\Tr \rho_{123}=1$. We have
$$
\rho_{12}=\left[ \begin {array}{cccc}  0&0&0&0\\\noalign{\medskip}0& \frac{1}{2}& \frac{1}{2}&0\\
\noalign{\medskip}0& \frac{1}{2}& \frac{1}{2}&0\\\noalign{\medskip}0&0&0&0\end {array} \right],
\quad
\rho_{23}=\left[ \begin {array}{cccc}  \frac{1}{4}& 0& 0& 0\\\noalign{\medskip} 0& \frac{1}{4}& 0& 0\\
\noalign{\medskip} 0& 0& \frac{1}{4}& 0\\\noalign{\medskip} 0& 0& 0& \frac{1}{4}\end {array} \right],
\quad
\rho_2=\left[ \begin {array}{cc}  \frac{1}{2}& 0\\\noalign{\medskip} 0& \frac{1}{2}\end {array} \right].
$$
One can compute that
$$
S_q(\rho_{123})+S_q(\rho_2)=\frac{1}{q-1}\ler{1-2\ler{\frac{1}{2}}^q+1-2\ler{\frac{1}{2}}^q}
=\frac{1}{q-1}\ler{2-4\ler{\frac{1}{2}}^q}
$$
and
$$
S_q(\rho_{12})+S_q(\rho_{23})=S_q(\rho_{23})=\frac{1}{q-1}\ler{1-4\ler{\frac{1}{4}}^q}.
$$
By the inequality of geometric and arithmetic means, we have $2 \cdot 2^{1-q}<1+4^{1-q}$,
which immediately shows that $S_q(\rho_{123})+S_q(\rho_2) > S_q(\rho_{12})+S_q(\rho_{23})$. \qed

This is not so surprising, if we consider a bit more general example.

\begin{ex}
If $\rho_{12}$ is an \emph{entangled} pure state, $1<\mathrm{rank}(\rho_3)$ and $\rho_{123}=
\rho_{12} \otimes \rho_{3},$ then
$$
S_q(\rho_{123})+S_q(\rho_2) > S_q(\rho_{12})+S_q(\rho_{23})
$$
holds for every $1<q.$ Indeed, if we use (\ref{classprod}) we get
$$
S_q (\rho_{123})+S_q(\rho_{2})=S_q (\rho_{12})+S_q(\rho_3)+(1-q)S_q(\rho_{12})S_q(\rho_3)+S_q(\rho_{2})
=S_q(\rho_{2})+S_q(\rho_3)
$$
and
$$
S_q(\rho_{12})+S_q(\rho_{23})=S_q(\rho_{23})=S_q(\rho_{2})+S_q(\rho_3)+(1-q)S_q(\rho_{2})S_q(\rho_3),
$$
because $S_q(\rho_{12})=0.$
$\rho_{12}$ is entangled, hence $S_q(\rho_{2})>0,$ and this verifies the statement.
\end{ex}

However, for classical probability distributions, Tsallis entropy is strongly
subadditive for $1 \leq q$ \cite{Furu}, and this result has a elegant and short
proof. The only thing necessary for the proof is that for any positive $x,y$ and $q,$ the identity
\be \label{comp_1}
\ln_{q}\, x -\ln_{q}\,y= -\ln_{q}\ler{\frac{y}{x}}x^{q-1}
\ee
holds.
Now we restate the proof of Furuichi.

\proof
If $\rho_{123}=\mathrm{Diag}(\{p_{jkl}\}),$ then by (\ref{comp_1}),
$$
S_q(\rho_{123})-S_q(\rho_{12})=-\sum_{j,k,l} p_{jkl} (\ln_q \ler{p_{jkl}} -\ln_q \ler{p_{jk}})
=\sum_{j,k,l} p_{jkl}^q \ln_q \ler{\frac{p_{jk}}{p_{jkl}}}
$$
$$
=\sum_{j,k,l} p_{jk}^q \ler{\frac{p_{jkl}}{p_{jk}}}^q \ln_q \ler{\frac{p_{jk}}{p_{jkl}}}.
$$
Observe that $x^q \ln_q \ler{\frac{1}{x}} =\Ln_q \, x,$ hence this expression can be written as
$$
\sum_{j,k,l} p_{jk}^q \Ln_q \ler{\frac{p_{jkl}}{p_{jk}}}=\sum_{j,k,l} \ler{\frac{p_{jk}}{p_k}}^q p_k^q
\Ln_q \ler{\frac{p_{jkl}}{p_{jk}}}\leq \sum_{k,l} p_k^q \ler{\sum_j \frac{p_{jk}}{p_k}  \Ln_q
\ler{\frac{p_{jkl}}{p_{jk}}}}.
$$
$\Ln_q$ is concave, hence
$$
\sum_{k,l} p_k^q \ler{\sum_j \frac{p_{jk}}{p_k}  \Ln_q \ler{\frac{p_{jkl}}{p_{jk}}}}
\leq \sum_{k,l} p_k^q \Ln_q\ler{\sum_j \frac{p_{jk}}{p_k}   \frac{p_{jkl}}{p_{jk}}}
=\sum_{k,l} p_k^q \Ln_q\ler{ \frac{p_{kl}}{p_k}}=
$$
$$
=\sum_{k,l} p_k^q \ler{\frac{p_{kl}}{p_{k}}}^q \ln_q \ler{\frac{p_{k}}{p_{kl}}}
=\sum_{k,l} p_{kl}^q \ln_q\ler{\frac{p_k}{p_{kl}}}=
-\sum_{k,l} p_{kl} (\ln_q \ler{p_{kl}} -\ln_q \ler{p_{k}})
$$
$$
=S_q(\rho_{23})-S_q(\rho_{2}).
$$
\qed


\section{Relative entropy and monotonicity}

Let $f$ be a $(0, \infty) \rightarrow \R$ function, let $\rho, \sigma \in \Mn$ be positive
definite matrices and $A \in \Mn$. We define the relative quasi-entropy by
\be \label{quasidef}
S_{f}^{A}\ler{\rho \, || \, \sigma}:=\inner{A \rho^{\frac{1}{2}}}{f\ler{\Delta(\sigma/\rho)}
\ler{A \rho^{\frac{1}{2}}}},
\ee
where $\inner{A}{B}= \tr A^* B$ is the Hilbert-Schmidt inner product and $\Delta(\sigma/\rho)$
is the relative modular operator introduced by Araki \cite{Araki}:
$$
\Delta(\sigma/\rho): \Mn \rightarrow \Mn\, \, , \quad  X \mapsto \sigma X \rho^{-1}.
$$
If $A=I,$ we simply write $S_{f}\ler{\rho || \sigma}.$

Whenever an expression $S_{f}^{A}\ler{\rho\, ||\, \sigma}$ appears, we implicitly assume that
$\rho$ is invertible. The following statement appeared in \cite{Sharma} and it makes
the relative quasi-entropy easy to compute in some cases.

\begin{lemma} \label{sharma}
Let the spectral decomposition of the positive definite matrices $\rho$ and $\sigma$
be given by
$$
\rho=\sum_j \lambda_j \ket{\varphi_j}\bra{\varphi_j} \text{ and } \sigma=
\sum_k \mu_k \ket{\psi_k}\bra{\psi_k}.
$$
Then we have
\be \label{relent-szamol}
S_{f}^{A}\ler{\rho \, || \, \sigma}=
\sum_{j,k} \lambda_j f\left(\frac{\mu_k}{\lambda_j}\right)\abs{ \bra{\psi_k} A \ket{\varphi_j}}^2.
\ee
\end{lemma}

\proof
$\left\{\ket{\psi_k} \bra{\varphi_j}\right\}_{j,k=1}^{n}$ form an orthonormal basis of
$\Mn$ (with respect to the Hilbert-Schmidt inner product.) It is easy to check that
with the notation $v_{jk}:=\ket{\psi_k} \bra{\varphi_j}$ we can write the relative
modular operator as
\be \label{relmodszam}
\Delta(\sigma/\rho)=\sum_{j,k} \frac{\mu_k}{\lambda_j} \ket{v_{jk}}\bra{v_{jk}},
\ee
where $ \ket{v_{jk}}\bra{v_{jk}}: \Mn \rightarrow \Mn$ is defined by $\ket{v_{jk}}
\bra{v_{jk}} (X):=v_{jk} \tr\, v_{jk}^{*} X.$ Therefore, $f\ler{\Delta(\sigma/\rho)}=
\sum_{j,k} f\ler{\frac{\mu_k}{\lambda_j}} \ket{v_{jk}}\bra{v_{jk}}.$ Direct computation shows that
$$
\inner{A \rho^{\frac{1}{2}}}{\ket{v_{jk}}\bra{v_{jk}}\ler{A  \rho^{\frac{1}{2}}}}=
$$
$$
=\tr \ler{\sum_a \lambda_a^{\frac{1}{2}} \ket{\varphi_a}\bra{\varphi_a} A^* \ket{\psi_k} \bra{\varphi_j}
\tr \ler{\ket{\varphi_j}\bra{\psi_k} A \sum_b \lambda_b^{\frac{1}{2}} \ket{\varphi_b}\bra{\varphi_b}}}
$$ $$
= \sum_{a,b}\lambda_a^{\frac{1}{2}} \lambda_b^{\frac{1}{2}}
\tr \ler{ \ket{\varphi_a}\bra{\varphi_a} A^* \ket{\psi_k} \bra{\varphi_j} \tr
\ler{\braket{\varphi_b}{\varphi_j}\bra{\psi_k} A  \ket{\varphi_b}}}
$$
$$
=\sum_{a,b} \lambda_a^{\frac{1}{2}} \lambda_b^{\frac{1}{2}} \delta_{bj} \bra{\psi_k} A \ket{\varphi_b}
\tr \ler{ \ket{\varphi_a}\bra{\varphi_a} A^* \ket{\psi_k} \bra{\varphi_j}}
$$
$$
= \sum_{a,b} \lambda_a^{\frac{1}{2}} \lambda_b^{\frac{1}{2}} \delta_{bj} \delta_{aj} \bra{\psi_k} A
\ket{\varphi_b} \bra{\varphi_a} A^* \ket{\psi_k}
=\lambda_j \abs{ \bra{\psi_k} A \ket{\varphi_j}}^2,
$$
therefore
$$
S_{f}^{A}\ler{\rho \,||\, \sigma}=\inner{A \rho^{\frac{1}{2}}}{\sum_{j,k} f\ler{\frac{\mu_k}{\lambda_j}}
\ket{v_{jk}}\bra{v_{jk}} \ler{A  \rho^{\frac{1}{2}}}}=\sum_{j,k} \lambda_j f\ler{\frac{\mu_k}{\lambda_j}}
\abs{ \bra{\psi_k} A \ket{\varphi_j}}^2.
$$
\qed

A short and elegant proof of the monotonicity of the relative entropy is given by Nielsen and
Petz in \cite{N-P}. The statement is that if $A, B \in \Mm \otimes \Mn$ are positive definite
matrices and we set $A_1=\tr_2 A, \ B_1=\tr_2 B,$ then for an operator convex function $f$ the
following inequality holds:

\be \label{np-mon}
S_{f}\ler{A ||B} \geq S_{f}\ler{A_1 ||B_1}.
\ee

There is a useful extension of this monotonicity in \cite{Sharma}. Now we restate Sharma's
proof, which is essentially the same as the proof of Nielsen and Petz \cite{N-P}.

\begin{lemma} \label{sufflem}
Suppose that $A, B \in \Mm \otimes \Mn$ are positive definite matrices and let us use the
notations $A_1=\tr_2 A, \ B_1=\tr_2 B.$ If $f$ is an operator convex function then for any
$T \in \Mm$ the following inequality holds:
\be \label{monext}
S_{f}^{T \otimes I_2}\ler{A ||B} \geq S_{f}^{T}\ler{A_1 ||B_1},
\ee
where $I_2$ is the identity in $\Mn.$
\end{lemma}

\proof
Let us consider the linear map
$$
\mc{U}: \Mm \rightarrow \Mm \otimes \Mn; \ X \mapsto \mc{U}(X):=
\ler{X A_1^{-\frac{1}{2}} \otimes I_2} A^{\frac{1}{2}}.
$$
We can check that $\mc{U}$ is an isometry. For $X,Y \in \Mm,$
$$
\inner{\mc{U}(X)}{\mc{U}(Y)}=\tr \ler{A^{\frac{1}{2}} \ler{ A_1^{-\frac{1}{2}}X^*
\otimes I_2} \ler{Y A_1^{-\frac{1}{2}} \otimes I_2} A^{\frac{1}{2}}}=
\tr \ler{A \ler{ A_1^{-\frac{1}{2}}X^* Y A_1^{-\frac{1}{2}} \otimes I_2}}
$$
$$
=\tr A_1  \ler{ A_1^{-\frac{1}{2}}X^* Y A_1^{-\frac{1}{2}}}=\tr X^* Y=\inner{X}{Y}.
$$
The short computation
\begin{eqnarray*}
\inner{Y}{\mc{U}(X)}& = &\tr Y^*\ler{X A_1^{-\frac{1}{2}} \otimes I_2} A^{\frac{1}{2}} =
\tr   \ler{YA^{\frac{1}{2}}}^* \ler{X  \otimes I_2} \ler{A_1^{-\frac{1}{2}} \otimes I_2}
\cr &=&
\tr \ler{A_1^{-\frac{1}{2}} \otimes I_2} \ler{YA^{\frac{1}{2}}}^* \ler{X  \otimes I_2} =
\tr  \ler{YA^{\frac{1}{2}} \ler{A_1^{-\frac{1}{2}} \otimes I_2}}^* \ler{X  \otimes I_2}
\cr &=&
\tr \ler{\tr_2 \ler{Y A^{\frac{1}{2}}\ler{A_1^{-\frac{1}{2}}\otimes I_2}}}^*X=
\inner{\tr_2 \ler{Y A^{\frac{1}{2}}\ler{A_1^{-\frac{1}{2}}\otimes I_2}}}{X}
\end{eqnarray*}
shows that the adjoint of $\mc{U}$ (which will be denoted by $\mc{U}^*$) is the map
$$
Y \mapsto \tr_2 \ler{Y A^{\frac{1}{2}}\ler{A_1^{-\frac{1}{2}}\otimes I_2}}.
$$
One can see that $\mc{U}$ admits the beautiful relation
\be \label{szep}
\mc{U}^* \Delta \ler{B/A} \mc{U} = \Delta \ler{B_1/A_1}.
\ee
If $X \in \Mm,$ then
\begin{eqnarray*}
\mc{U}^* \Delta \ler{B/A} \mc{U}(X) & = &  \tr_2 \ler{B \ler{X A_1^{-\frac{1}{2}} \otimes I_2}
A^{\frac{1}{2}} A^{-1} A^{\frac{1}{2}}\ler{A_1^{-\frac{1}{2}}\otimes I_2}}
\cr &=&
\tr_2 \ler{B\ler{X A_1^{-1} \otimes I_2}}=B_1 X A_1^{-1}=\Delta \ler{B_1/A_1}(X).
\end{eqnarray*}

By definition of the relative entropy and by (\ref{szep}), the right-hand-side of (\ref{monext})
can be written as
$$
S_{f}^{T}\ler{A_1 ||B_1}=\inner{T A_1^{\frac{1}{2}}}{f\ler{\Delta(B_1/A_1)} \ler{T A_1^{\frac{1}{2}}}}=
\inner{T A_1^{\frac{1}{2}}}{f\ler{\mc{U}^* \Delta \ler{B/A} \mc{U}} \ler{T A_1^{\frac{1}{2}}}}.
$$
The operator convexity of $f$ implies that
$$f\ler{\mc{U}^* \Delta \ler{B/A} \mc{U}} \leq \mc{U}^* f\ler{\Delta \ler{B/A}} \mc{U}$$
(see Chapter 5 of \cite{Bhat}), and
$\mc{U}\ler{T A_1^{\frac{1}{2}}}=\ler{T \otimes I_2}A^{\frac{1}{2}}$ is immediate. Therefore,
$$
\inner{T A_1^{\frac{1}{2}}}{f\ler{\mc{U}^* \Delta \ler{B/A} \mc{U}} \ler{T A_1^{\frac{1}{2}}}} \leq
\inner{\mc{U}\ler{T A_1^{\frac{1}{2}}}}{f\ler{\Delta \ler{B/A}} \ler{\mc{U}\ler{T A_1^{\frac{1}{2}}}}}
$$
$$
=\inner{\ler{T \otimes I_2}A^{\frac{1}{2}}}{f\ler{\Delta \ler{B/A}}
\ler{\ler{T \otimes I_2}A^{\frac{1}{2}}}}=S_{f}^{T \otimes I_2}\ler{A ||B},
$$
and the proof is complete. \qed


\section{From the relative entropy to the strong subadditivity}

As we have seen in Section \ref{bev}, the strong subadditivity of the Tsallis entropy holds if
and only if $q=1.$ Therefore, our goal is to find some formulas as
\begin{equation}\label{Tub14}
S_q(\rho_{123})+S_q(\rho_2)\leq S_q(\rho_{12})+S_q(\rho_{23})+f_q(\rho_{123}),
\end{equation}
where $f_1(\rho_{123})=0$. Such a result can be considered as a generalization of the SSA inequality.

Now we collect here some elementary facts that will be used in this section.
\begin{enumerate}
\item
For any positive $x,y$ and $q,$ the identity
\be \label{comp_2}
\ln_{q}\, x-\ln_{q}\,y = -\ln_{q}\ler{\frac{y}{x}}-(q-1)\ln_{q}\ler{\frac{y}{x}} \ln_q\,x
\ee
holds.

\item
If $f$ and $g$ are $\R \rightarrow \R$ functions, $\rho, \sigma \in \Mnsa$ and the spectral
decompositions are $\rho=\sum_j \lambda_j \ket{\varphi_j}\bra{\varphi_j}$  and  $\sigma=\sum_k
\mu_k \ket{\psi_k}\bra{\psi_k},$ and domain of $f$ and $g$ contains the spectrum of $\rho$ and
$\sigma,$ respectively, then
\be \label{fcalc}
\tr f(\rho) g(\sigma)=\sum_{j,k} f(\lambda_j) g(\mu_k)\abs{ \braket{\varphi_j}{\psi_k}}^2.
\ee

\item
If $A \in \Mm \otimes \Mn$ and $B \in \Mm,$ then
\be  \label{part}
\tr_2 \ler{A \cdot B \otimes I_2}=(\tr_2 A) \cdot B.
\ee
\end{enumerate}

The strong subadditivity of the von Neumnann entropy can be derived from the monotonicity of
the  Umegaki relative entropy \cite{Carlen, N-P}. Therefore, it seems to be useful
to reformulate the SSA of the Tsallis entropy as an inequality of relative quasi-entropies.
\begin{Thm} \label{basic_1}
The strong subadditivity inequality of the Tsallis entropy (\ref{Tub12}) is equivalent to
\be \label{thm1}
S_{-\ln_q}^U \ler{\rho_{123}\, ||\,\rho_{12} \otimes I_3} \geq
S_{-\ln_q}^V \ler{\rho_{23}\, ||\, \rho_{2} \otimes I_3},
\ee
where
\be \label{thm11}
U=\rho_{123}^{\frac{1}{2}(q-1)}, \qquad V=\rho_{23}^{\frac{1}{2}(q-1)}\,.
\ee
\end{Thm}

\proof
By (\ref{part}),
$$
\tr_3 \ler{\rho_{123} \ln_q(\rho_{12} \otimes I_3)}=\rho_{12}  \ln_q(\rho_{12})
\text { and } \tr_3 \ler{\rho_{23} \ln_q(\rho_{2} \otimes I_3)}=\rho_{2}  \ln_q(\rho_{2}),
$$
hence the inequality
$$
-S_q(\rho_{123})+S_q(\rho_{12}) \geq -S_q(\rho_{12})+S_q(\rho_2),
$$
which is obviously equivalent to (\ref{Tub12}), can be written in the form
\be \label{ssa2}
\tr \rho_{123} \ler{\ln_q(\rho_{123})-\ln_q(\rho_{12} \otimes I_3)} \geq \tr \rho_{23}
\ler{\ln_q(\rho_{23})-\ln_q(\rho_{2} \otimes I_3)}.
\ee

By (\ref{fcalc}), if $\rho_{123}=\sum_j \lambda_j \ket{\varphi_j}\bra{\varphi_j}$
and  $\rho_{12} \otimes I_3=\sum_k \mu_k \ket{\psi_k}\bra{\psi_k},$ then the left hand side
of (\ref{ssa2}) is
$$
\sum_{j,k} \lambda_j \ler{\ln_q{\lambda_j}-\ln_q{\mu_k}}\abs{ \braket{\varphi_j}{\psi_k}}^2.
$$

By (\ref{comp_1}), this expression can be written as
\be \label{entfele_1}
\sum_{j,k} \lambda_j \ler{ -\ln_{q}\ler{\frac{\mu_k}{\lambda_j}}\lambda_j^{q-1}}
\abs{ \braket{\varphi_j}{\psi_k}}^2.
\ee

In addition, if $U=\rho_{123}^{\frac{1}{2}(q-1)}$ then
$$
\abs{ \bra{\psi_k} U\ket{\varphi_j}}^2=\lambda_j^{q-1}\abs{ \braket{\psi_k}
{\varphi_j}}^2,
$$
hence by the result of Lemma \ref{sharma}, we can write (\ref{entfele_1}) as the following
relative entropy
\be \label{atir_1}
\sum_{j,k} \lambda_j \ler{ -\ln_{q}\ler{\frac{\mu_k}{\lambda_j}}}\abs{ \bra{\psi_k}
U \ket{\varphi_j}}^2=
S_{-\ln_q}^{U}\ler{\rho_{123} \, ||\, \rho_{12} \otimes I_3}.
\ee
The observation that the right hand side of (\ref{ssa2}) can be written as a relative
entropy similarly to (\ref{atir_1}) completes the proof.
\qed

Note that in the special case $q=1,$ Theorem \ref{basic_1} states the equivalence of
the monotonicity of the Umegaki relative entropy and the SSA of the von Neumann entropy.

\begin{Thm} \label{lim}
For $0 < q \leq 2$ the inequality
$$
S_q(\rho_{12})+S_q(\rho_{23}) -S_q(\rho_{123})-S_q(\rho_{2})
$$
$$
\geq
(q-1)\ler{S_{\ln_q}^{\ler{-\ln_q\rho_{123}}^{\frac{1}{2}}}\ler{\rho_{123} ||\rho_{12} \otimes I_3}
-S_{\ln_q}^{\ler{-\ln_q\rho_{23}}^{\frac{1}{2}}}\ler{\rho_{23} \,||\, \rho_{2} \otimes I_3}}
$$
holds.
\end{Thm}

Remark that this statemant recovers the strong subadditivity of the von Neumann entropy if $q=1$.

\proof
We noted that if $\rho_{123}=\sum_j \lambda_j \ket{\varphi_j}\bra{\varphi_j}$
and  $\rho_{12} \otimes I_3=\sum_k \mu_k \ket{\psi_k}\bra{\psi_k},$ then the left hand side
of (\ref{ssa2}) is
$$
\sum_{j,k} \lambda_j \ler{\ln_q{\lambda_j}-\ln_q{\mu_k}}\abs{ \braket{\varphi_j}{\psi_k}}^2.
$$
According to (\ref{comp_2}), it is equal to
\be \label{entfele_2}
\sum_{j,k} \lambda_j \ler{ -\ln_{q}\ler{\frac{\mu_k}{\lambda_j}}-(q-1)\ln_{q}\ler{\frac{\mu_k}
{\lambda_j}} \ln_{q} \lambda_j}\abs{ \braket{\varphi_j}{\psi_k}}^2,
\ee
and by Lemma \ref{sharma}, (\ref{entfele_2}) has the form
\be \label{atir_2}
S_{-\ln_q}\ler{\rho_{123}\, ||\, \rho_{12} \otimes I_3}+(q-1)S_{\ln_q}^{\ler{-\ln_q\rho_{123}}^{\frac{1}{2}}}
\ler{\rho_{123} \,||\, \rho_{12} \otimes I_3}.
\ee

If we rewrite the right hand side of (\ref{ssa2}) similarly to (\ref{atir_2}), we get that the
SSA is equivalent to
$$
S_{-\ln_q}\ler{\rho_{123} \,||\,\rho_{12} \otimes I_3}+(q-1)S_{\ln_q}^{\ler{-\ln_q\rho_{123}}^{\frac{1}{2}}}
\ler{\rho_{123} \, ||\, \rho_{12} \otimes I_3}
$$
\be \label{kul}
\geq
S_{-\ln_q}\ler{\rho_{23} \,||\, \rho_{2} \otimes I_3}+(q-1)S_{\ln_q}^{\ler{-\ln_q\rho_{23}}^{\frac{1}{2}}}
\ler{\rho_{23} \,||\, \rho_{2} \otimes I_3}.
\ee

It is easy to derive from the L\"owner-Heinz theorem \cite{Bhat, Carlen, H-P} that
$\ln_{q} \, x$ is operator monotone, if $0 < q \leq 2.$ An operator monotone function is
operator concave \cite{H-P}, hence $-\ln_{q}\,x $ is operator convex.

By the monotonicity property (\ref{monext}), for $0 < q \leq 2$ we have
$$
S_{-\ln_q}\ler{\rho_{123}\, ||\, \rho_{12} \otimes I_3} \geq
S_{-\ln_q}\ler{\rho_{23} \, ||\, \rho_{2} \otimes I_3},
$$
and by (\ref{kul}), this is equivalent to
$$
-S_q(\rho_{123})+S_q(\rho_{12})-(q-1)\ler{S_{\ln_q}^{\ler{-\ln_q\rho_{123}}^{\frac{1}{2}}}
\ler{\rho_{123} \, ||\, \rho_{12} \otimes I_3}}
$$
\be \label{ossz}
\geq
-S_q(\rho_{23})+S_q(\rho_{2})-(q-1)\ler{S_{\ln_q}^{\ler{-\ln_q\rho_{23}}^{\frac{1}{2}}}
\ler{\rho_{23} \, ||\, \rho_{2} \otimes I_3}}.
\ee
This is the statement of the theorem.
\qed

The notation (\ref{thm11}) will be used again. Because of the monotonicity property
(\ref{monext}), for $0<q \leq 2$ and $f(x)=-\ln_q\, x, A=\rho_{123}, B=\rho_{12}\otimes I_3,
T=V$ we have
\be \label{sh-n-p}
S_{-\ln_q}^{I_1 \otimes V}\ler{\rho_{123} \,||\, \rho_{12} \otimes I_3} \geq
S_{-\ln_q}^{V}\ler{\rho_{23} \,||\,\rho_{2} \otimes I_3}.
\ee
This formula is quite similar to the SSA inequality (\ref{thm1}). By (\ref{sh-n-p}),
\be \label{jolenne}
S_{-\ln_q}^{U}\ler{\rho_{123} \,||\,\rho_{12} \otimes I_3}
\geq S_{-\ln_q}^{I_1 \otimes V}\ler{\rho_{123} ||\rho_{12} \otimes I_3}
\ee
implies the strong subadditivity (\ref{thm1}).  We try to find a sufficient condition
for (\ref{jolenne}).

\begin{Thm} \label{suff}
If $\rho_{123}$ and $I_1 \otimes \rho_{23}$ commute, and (using the usual notation
$\rho_{123}=\sum_j \lambda_j \ket{\varphi_j}\bra{\varphi_j}$  and  $\rho_{12} \otimes I_3
=\sum_k \mu_k \ket{\psi_k}\bra{\psi_k}$) we have $\lambda_j \leq \mu_k$ whenever
$\braket{\psi_k}{\varphi_j} \neq 0$, then for any $1 \leq q \leq 2$ the strong
subadditivity inequality
$$
S_q(\rho_{123})+S_q(\rho_{2}) \leq S_q(\rho_{12})+S_q(\rho_{23})
$$
holds.
\end{Thm}

Note that if $\rho_{123}$ is a classical probability distribution (that is, it is diagonal
in a product basis), then the conditions of Theorem \ref{suff} are clearly satisfied.

\begin{proof}
By Lemma \ref{sharma}, (\ref{jolenne}) is equivalent to
$$ \label{jolenne2}
\sum_{j,k} \abs{\bra{\psi_k} \rho_{123}^{\frac{1}{2}(q-1)} \ket{\varphi_j}}^2
\ler{-\ln_q \ler{\frac{\mu_k}{\lambda_j}}} \lambda_j
\geq
\sum_{j,k} \abs{\bra{\psi_k} I_1 \otimes \rho_{23}^{\frac{1}{2}(q-1)} \ket{\varphi_j}}^2
\ler{-\ln_q \ler{\frac{\mu_k}{\lambda_j}}} \lambda_j\, .
$$
If $\lambda_j \leq \mu_k$, then $ \ler{-\ln_q \ler{\frac{\mu_k}{\lambda_j}}} \lambda_j \leq 0.$
On the other hand,
$$
\abs{\bra{\psi_k} \rho_{123}^{\frac{1}{2}(q-1)} \ket{\varphi_j}}^2=\lambda_j^{q-1}
\abs{\braket{\psi_k}{\varphi_j}}^2.
$$
If $\rho_{123}$ and $I_1 \otimes \rho_{23}$ commute, then $I_1 \otimes \rho_{23}$ is
diagonal in the basis $\left\{\varphi_j\right\}_{j \in J}$ and $\rho_{123} \leq I_1 \otimes
\rho_{23}$ holds, that is, $I_1 \otimes \rho_{23}=\sum_j \nu_j \ket{\varphi_j}\bra{\varphi_j}$
with some $\nu_j \geq \lambda_j.$

If $1 \leq q,$ the map $t \mapsto t^{(q-1)}$ is monotone on $\R_{+}$, hence we have
$$
\abs{\bra{\psi_k} I_1 \otimes \rho_{23}^{\frac{1}{2}(q-1)} \ket{\varphi_j}}^2=
\nu_j^{q-1} \abs{\braket{\psi_k}{\varphi_j}}^2 \geq \lambda_j^{q-1} \abs{\braket{\psi_k}{\varphi_j}}^2.
$$
We concluded that if the conditions of Theorem \ref{suff} are satisfied, then (\ref{jolenne})
holds, and hence the proof is complete.
\qed
\end{proof}
The following example shows that one can apply Theorem \ref{suff} in essentially
non-classical cases, as well.

\begin{ex}
Set $p, q \in \left[\frac{1}{2} ,1\right]$ such that $p q \leq 1-q$ and
$t \in \R.$ Let us define $V$ and $\Lambda$ by
$$
V=
\left[
\begin{array}{cccc}
 \cos{t} & 0 & 0 & -\sin{t} \\
 0 & \cos{t} & -\sin{t} & 0 \\
 0 & \sin{t} & \cos{t} & 0 \\
 \sin{t} & 0 & 0 & \cos{t} \\
\end{array}
\right], \, \Lambda=\mathrm{Diag}(p q, (1-p)q, p (1-q), (1-p)(1-q)).
$$
V describes a family of orthonormal bases, this family can be considered as a one-parameter
extension of the Bell basis. Let $\rho_1 \in \Mm$ be an arbitrary density, and $\rho_{23}
\in M_{2}(\C) \otimes M_2(\C)$ be defined by
$$
\rho_{23}=V \Lambda V^{-1}=
\left[\begin{array}{cccc}
p q \cos^{2} t +(1-p)(1-q) \sin^2 t & 0 & 0 & (p q -(1-p)(1-q))\sin t \cos t \\
 0 & A_{11} & A_{12} & 0 \\
 0 & A_{21} & A_{22} & 0 \\
 (p q -(1-p)(1-q))\sin t \cos t & 0 & 0 &  (1-p)(1-q)\cos^{2} t + p q \sin^2 t
\end{array}\right],
$$
where
$$
A_{11}=(1-p) q \cos^{2} t +p(1-q) \sin^2 t\, , \qquad A_{12}= ((1-p)q -p(1-q))\sin t \cos t\, ,
$$ $$
A_{21}=((1-p)q -p(1-q))\sin t \cos t \,, \qquad A_{22}= p(1-q)\cos^{2} t + (1-p) q \sin^2 t \,.
$$

One can easily compute that
$$
\rho_2 =\tr_3 \,\rho_{23}=
\left[
\begin{array}{cc}
q \cos^2 t+(1-q) \sin^2 t & 0 \\
0 & q \sin^2 t + (1-q) \cos^2 t
\end{array}\right].
$$
Let us take the density $\rho_{123}=\rho_1 \otimes \rho_{23}.$ The spectrum of $\rho_{123}$ is
$$
\bigcup_{j=1}^{m} \{\nu_j p q, \nu_j (1-p)q,\nu_j p (1-q),\nu_j (1-p) (1-q)\},
$$
where $\nu_j$'s are the eigenvalues of $\rho_1$. The spectrum of $\rho_{12}$
(or $\rho_{12} \otimes I_3$) is
$$
\bigcup_{j=1}^{m} \{\nu_j (q \cos^2 t+(1-q) \sin^2 t), \nu_j (q \sin^2 t + (1-q) \cos^2 t)\}.
$$
The assumption $p q \leq 1-q$ guarantees that the eigenvalues of $\rho_{123}$ are smaller
than the eigenvalues of $\rho_{12} \otimes I_3$, whenever the corresponding eigenvectors are
not orthogonal. $\rho_{123}$ and $I_1 \otimes \rho_{23}$ obviously commute, hence the conditions
of Theorem \ref{suff} are satisfied by $\rho_{123}$ despite the fact that $\rho_{23}$ can not
be diagonized in any product basis.
\end{ex}

{\bf Acknowledgments.} This work was partially supported by the Hungarian Research Grant
OTKA K104206. The authors would like to thank the referee for providing constructive comments.


\end{document}